\documentclass[journal, final]{IEEEtran}
\usepackage{cite}
\usepackage{amsmath,amssymb,amsfonts}
\usepackage{algorithm}
\usepackage{algorithmic}
\usepackage{graphicx}
\usepackage{textcomp}
\usepackage{bm}
\usepackage{booktabs}
\usepackage{fancyhdr}
\usepackage{geometry}
\def\BibTeX{{\rm B\kern-.05em{\sc i\kern-.025em b}\kern-.08em
    T\kern-.1667em\lower.7ex\hbox{E}\kern-.125emX}}

\begin{document}

\title{Convolutional Sparse Coding for Compressed Sensing CT Reconstruction}

\author{Peng Bao, Wenjun Xia, Kang Yang, Weiyan Chen, Mianyi Chen, Yan Xi, Shanzhou Niu, Jiliu Zhou, \IEEEmembership{Senior Member, IEEE}, He Zhang, Huaiqiang Sun, Zhangyang Wang, and Yi Zhang, \IEEEmembership{Senior Member, IEEE}
\thanks{This work was supported in part by the National Natural Science Foundation of China under grants 61671312, the Sichuan Science and Technology Program under grants 2018HH0070 and Miaozi Project in Science and Technology Innovation Program of Sichuan Province under grants 18-YCG041. (Corresponding author: Yi Zhang.)}
\thanks{P. Bao, W. Xia, K. Yang, J. Zhou and Y. Zhang are with the College of Computer Science, Sichuan University, Chengdu 610065, China (email: pengbao7598@gmail.com; xwj90620@gmail.com; 18811352562@163.com; zhoujl@scu.edu.cn; ;yzhang@scu.edu.cn).}
\thanks{W. Chen is with the Key Laboratory of High Confidence Software Technologies, Ministry of Education, Beijing, China and the School of Electronics Engineering and Computer Science, Peking University, China (email: chenwy1117@pku.edu.cn).}
\thanks{M. Chen and Y. Xi are with Shanghai First-imaging Information Technology co. LTD (email: mianyichen@first-imaging.com; yanxi@first-imaging.com).}
\thanks{S. Niu is with the school of Mathematics and Computer Science, Gannan Normal University, Ganzhou 341000, China (email: shanzhou.niu@gmail.com).}
\thanks{H. Zhang is with Rutgers, the State University of New Jersey, NJ, 08854 (email: zhanghe920312@gmail.com).}
\thanks{H. Sun is with Department of Radiology, West China Hospital of Sichuan University, Chengdu 610041, China (email: sunhuaiqiang@scu.edu.cn).}
\thanks{Z. Wang is with the Department of Computer Science and Engineering, Texas A\&M University, United States (email: atlaswang@tamu.edu).}
}

\maketitle
\thispagestyle{fancy} 
\cfoot{\copyright 2019 IEEE. Personal use of this material is permitted. Permission from IEEE must be obtained for all other uses, including reprinting/republishing this material for advertising or promotional purposes, collecting new collected works for resale or redistribution to servers or lists, or reuse of any copyrighted component of this work in other works.} 

\begin{abstract}
Over the past few years, dictionary learning (DL)-based methods have been successfully used in various image reconstruction problems. However, traditional DL-based computed tomography (CT) reconstruction methods are patch-based and ignore the consistency of pixels in overlapped patches. In addition, the features learned by these methods always contain shifted versions of the same features. In recent years, convolutional sparse coding (CSC) has been developed to address these problems. In this paper, inspired by several successful applications of CSC in the field of signal processing, we explore the potential of CSC in sparse-view CT reconstruction. By directly working on the whole image, without the necessity of dividing the image into overlapped patches in DL-based methods, the proposed methods can maintain more details and avoid artifacts caused by patch aggregation. With predetermined filters, an alternating scheme is developed to optimize the objective function. Extensive experiments with simulated and real CT data were performed to validate the effectiveness of the proposed methods. Qualitative and quantitative results demonstrate that the proposed methods achieve better performance than several existing state-of-the-art methods.
\end{abstract}


\section{INTRODUCTION}
\label{sec:intro}
\IEEEPARstart{O}{ver} the past few decades, computed tomography (CT) has been very successfully used for clinical diagnosis and intervention. However, X-rays can be absorbed partially by the human body and may cause genetic or cancerous diseases \cite{Brenner2007Computed}. As a result, the famous ALARA (as low as reasonably achievable) principle has been widely utilized to avoid an excessive radiation dose in clinical practice. Reducing the X-ray flux towards each detector element and decreasing the number of sampling views are two effective ways to reduce the radiation dose, resulting in noisy measurements or insufficient projection data, respectively. However, it is chanllenging to generate high-quality images with degraded projection data using traditional methods, such as filtered back projection (FBP). Thus, reconstructing high-quality images from contaminated and/or undersampled projection data has attracted increased attention in the field of CT imaging. In this work, we focus on the sparse-view problem.

To recover from the situation of undersampling, iterative reconstruction methods have been developed, which can efficiently improve the image quality when the projection views are incomplete. Classical iterative reconstruction algorithms, such as the algebraic reconstruction technique (ART) \cite{Gordon1970Algebraic}, simultaneous the algebraic reconstruction technique (SART) \cite{Andersen1984Simultaneous} and expectation maximization (EM) \cite{Dempster1977Maximum}, can mitigate this problem to a certain degree. However, it is difficult to achieve a satisfactory result when projection views are highly sparse without extra priors. Introducing reasonable priors can significantly improve the imaging quality of traditional iterative methods. In recent years, compressed sensing (CS) theory has been developed and has been proven to be a powerful tool, which indicates that a signal can be exactly reconstructed with a very high probability if the signal can be represented sparsely with a certain sparsifying transform \cite{Cand2006Robust, Donoho2006Compressed}. Inspired by CS theory, Sidky \emph{et al.} \cite{Sidky2006Accurate, Sidky2008Image} coupled total variation (TV) and projection onto convex sets (POCS) to solve incomplete projection data reconstruction problems and achieved good performance. However, TV assumes that the signal is piecewise smooth, resulting in undesired patchy effects on the reconstructed images. Following this path, many variants of TV have been proposed, such as adaptive-weighted TV \cite{Liu2012Adaptive}, fractional-order TV \cite{Zhang2013Few, Zhang2014Few}, and nonlocal TV \cite{Kim2016Non,Zhang2016Spectral}. Furthermore, Niu \emph{et al.} \cite{Niu2014Sparse} combined penalized weighted least-squares (PWLS) \cite{Fessler1994Penalized} with total generalized variation (TGV) \cite{Bredies2010Total} to reduce the patchy effects. Chen and Lang indicated that a high-quality image can be used as the prior to accurately reconstruct CT dynamic image sequences from highly undersampled projection data \cite{Chen2008Prior}. 

Recently, inspired by the rapid development of deep learning, several pioneering studies combining the ideas of deep learning and CS were presented \cite{Wang2016A, Wang2018Image}. Typically, Chen \emph{et al.} proposed to unfold the iterative reconstruction into a residual convolutional neural network, and the parameters and the regularization terms can be learned from the external dataset \cite{Chen2018LEARN}. Alder and {\"O}ktem employed the primal dual hybrid gradient (PDHG) algorithm for non-smooth convex optimization and unrolled the optimization process to a neural network \cite{Adler2018Learned}. Zhu \emph{et al.} proposed a unified image reconstruction framework--automated transform by manifold approximation (AUTOMAP)--which directly learned a potential mapping function from the sensor domain to the image domain \cite{Zhu2018Image}. In \cite{He2019Optimizing}, a parameterized plug-and-play alternating direction method of multipliers (3pADMM) was developed for the PWLS model, and then the prior terms and related parameters were optimized with a neural network with a large quantity of training data. Except for the reconstruction problem, deep learning was also used in low-dose CT (LDCT) denoising. As a pioneering work, Chen \emph{et al.} used a three-layer convolutional neural network (CNN) to handle the noise in the LDCT images \cite{Chen2017Low2}. Kang \emph{et al.} introduced contourlet-based multi-scale U-Net to improve the image quality of LDCT \cite{Kang2017Deep}. Motivated by the idea of autoencoders, Chen \emph{et al.} proposed a residual encoder-decoder CNN (RED-CNN) for LDCT \cite{Chen2017Low}. These methods obtained promising results, but the requirement of a large number of training samples may limit the application scenarios.

As another direction of CS, in recent years, dictionary learning (DL)-based methods have been proven useful in various image restoration problems. In 2006, Elad and Aharon adopted a dictionary-based method to address the image denoising problem \cite{Elad2006Image}. Mairal \emph{et al.} explored how to apply this method to color image restoration \cite{Mairal2008Sparse}. In the field of medical imaging, the DL method was first introduced into magnetic resonance imaging (MRI) by Chen \emph{et al.} \cite{Chen2010A}. Ravishankar and Bresler utilized an adaptive dictionary-based method to reconstruct MR images from the highly undersampled $k$-space data \cite{Ravishankar2011MR}. Later, Xu \emph{et al.} first introduced DL into LDCT reconstruction \cite{Xu2012Low}. Two types of dictionaries were learned as bases: (1) a global dictionary was learned from an external training set that was fixed during the reconstruction procedure, and (2) an adaptive dictionary was learned from the intermediate result that continued to update during the iterations. Inspired by the work of super-resolution, which uses two dictionaries to link low-resolution images and the corresponding high-resolution training images \cite{Yang2010Image}, Lu \emph{et al.} proposed using two dictionaries to predict full-sampled CT images \cite{Lu2012Few}, and Zhao \emph{et al.} extended this method to spectral CT \cite{Zhao2012Dual}. Chen \emph{et al.} further improved the DL-based artifact reduction method with tissue feature atoms and artifact atoms to build discriminative dictionaries \cite{Chen2014Artifact}. Liu \emph{et al.} extended this idea to 3D sinogram restoration for LDCT \cite{Liu2017Discriminative}. Zheng \emph{et al.} proposed to learn a union of sparsifying transforms instead of dictionary atoms for LDCT image reconstruction \cite{Zheng2018PWLS}.

However, most dictionary learning methods are patch-based, and the learned features often contain shifted versions of the same features. The learned dictionary may be over-redundant, and the results will suffer from patch aggregation artifacts on the patch boundaries. To address these problems, convolutional sparse coding (CSC) methods have been proposed in which the shift invariance is directly modeled in the objective function. CSC can be traced back to \cite{Lewicki1999Coding} and was first proposed by Zeiler \emph{et al.} \cite{Zeiler2010Deconvolutional}. The effectiveness of CSC has been demonstrated in several computer vision tasks. Gu \emph{et al.} introduced CSC into super-resolution and achieved promising results \cite{Gu2015Convolutional}. Wohlberg approached impulse noise via CSC aided by gradient constraints \cite{Wohlberg2016Convolutional}. Liu \emph{et al.} applied CSC for image fusion \cite{Liu2016Image}. More recently, Zhang and Patel combined CSC and low-rank decomposition to address rain streak removal and image decomposition problems \cite{Zhang2017Convolutional, Zhang2018Convolutional}. Thanh \emph{et al.}proposed to adopt a shared 3D multi-scale convolutional dictionary to recover the high-frequency information of the MRI images \cite{Thanh2019Frequency}.

Inspired by the pioneering studies of CSC, in this work, two models based on CSC are proposed to address the sparse-view CT reconstruction problem. In the first model, which is called PWLS-CSC, a CSC regularization is adapted with the PWLS model. The filters are trained with an external full-sampled CT dataset. Although PWLS-CSC can work well for sparse-view CT, ringing artifacts may appear in some reconstructed images due to inaccurate filters. On the other hand, gradient constraint is an effective way to suppress these types of artifacts. Therefore, in the second model, an improved version of PWLS-CSC, which imposes gradient regularization on feature maps, is proposed. For simplicity, we called this method PWLS-CSCGR.

The remainder of this paper is organized as follows. In Section \ref{sec:background}, we introduce the background on PWLS and CSC. In Section \ref{sec:method}, the proposed reconstruction schemes are elaborated. In Section \ref{sec:results}, the experimental designs and representative results are given. Finally, we will discuss relevant issues and conclude this paper in Section \ref{sec:discs and concs}.

\section{BACKGROUND}
\label{sec:background}

\subsection{Penalized Weighted Least-Squares for CT Image Reconstruction}
\label{subsec:PWLS}
The calibrated and log-transformed projection data approximately follow a Gaussian distribution, which can be described by the following analytic formula \cite{Li2004Nonlinear}:
\begin{equation}
  \label{equ:variance}
  \sigma_p^2 = r_p \times exp(\bar{\bm{y}}_p/\varepsilon)
\end{equation}
where $\bar{\bm{y}}_p$ and $\sigma_p^2$ are the mean and variance of the measured projection data at $p$-th bin, respectively, $r_p$ is a parameter adaptive to different bins and $\varepsilon$ is a scaling parameter. Based on these properties of the projection data, the PWLS-based CT reconstruction problem can be modeled as follows:
\begin{equation}
  \label{equ:pwls}
  \mathop{arg\ min}_{\bm{u}}(\bm{y}-\bm{Au})^T \bm{\Sigma}^{-1} (\bm{y}-\bm{Au}) + \beta R(\bm{u})
\end{equation}
where $\bm{y}$ denotes the projection data, $\bm{u}$ is the vectorized attenuation coefficients to be reconstructed and $(\cdot)^T$ denotes the transpose operation. $\bm{A}$ denotes the system matrix with a size of $P \times Q$ ($P$ is the total number of projection data and $Q$ is the total number of image pixels). $\bm{\Sigma}$ is a diagonal matrix with the $p$-th element of $\sigma_p^2$ as calculated by (\ref{equ:variance}). $R(\bm{u})$ represents the regularization term, and $\beta$ is a balancing parameter to control the tradeoff between the data-fidelity and regularization terms.

\subsection{Convolutional Sparse Coding}
In contrast to traditional patch-based DL methods, CSC assumes that the image can be represented as a summation of convolutions between a set of filters and its corresponding feature maps:
\begin{equation}
  \label{equ:csc}
  \mathop{arg\ min}_{\{\bm{M}_i\},\{\bm{f}_i\}}\frac{1}{2}\left\|\sum_{i=1}^N \bm{f}_i \ast \bm{M}_i - \bm{x}\right\|_2^2 + \lambda \sum_{i=1}^N\left\|\bm{M}_i\right\|_1,
\end{equation}
where $\bm{x}$ denotes the image, $\ast$ denotes the convolution operator, $\{\bm{f}_i\}_{i=1,2,\dots,N}$ is a set of filters, $\{\bm{M}_i\}_{i=1,2,\dots,N}$ are the feature maps corresponding to filters $\bm{f}_i$, and $\lambda$ is the regularization parameter. In the CSC model, the feature map $\bm{M}_i$ has the same size as image $\bm{x}$. Since the reconstructed image is obtained by the summation of convolution operations $\bm{f}_i \ast \bm{M}_i$, the patch aggregation artifacts in traditional patch-based sparse coding methods can be avoided. Meanwhile, the features learned by patch-based sparse coding methods often contain shifted versions of the same features, and the filters learned by CSC can obtain rotation-invariant features, which makes this method more efficient.

\section{METHOD}
\label{sec:method}

\subsection{PWLS-CSC Model}
\label{subsec:PWLS-CSC}
Current TV- or DL-based methods suffer from blocky effects or patch aggregation artifacts. To circumvent these problems, we propose introducing CSC as the regularization term in (\ref{equ:pwls}). Combining CSC with (\ref{equ:pwls}), we aim to solve the following optimization problem:
\begin{equation}
  \label{equ:pwls-csc-1}
  \begin{split}
  &\mathop{arg\ min}_{\bm{u},\{\bm{M}_i\},\{\bm{f}_i\}}
  \frac{1}{2}(\bm{y}-\bm{Au})^T \bm{\Sigma}^{-1} (\bm{y}-\bm{Au}) + \\ &\beta\left(\frac{1}{2}\left\|\sum_{i=1}^N \bm{f}_i \ast \bm{M}_i - \bm{u}\right\|_2^2 + \lambda \sum_{i=1}^N \left\| \bm{M}_i \right\|_1\right).
  \end{split}
\end{equation}
In this model, we use an external full-sampled CT dataset to train the filters $\{\bm{f}_i\}$ by the method proposed in \cite{Wohlberg2016Efficient}, so (\ref{equ:pwls-csc-1}) becomes the following optimization problem:
\begin{equation}
  \label{equ:pwls-csc-2}
  \begin{split}
  &\mathop{arg\ min}_{\bm{u},\{\bm{M}_i\}}
  \frac{1}{2}(\bm{y}-\bm{Au})^T \bm{\Sigma}^{-1} (\bm{y}-\bm{Au}) + \\
  &\beta\left(\frac{1}{2}\left\|\sum_{i=1}^N \bm{f}_i \ast \bm{M}_i - \bm{u}\right\|_2^2 + \lambda \sum_{i=1}^N \left\| \bm{M}_i \right\|_1\right).
  \end{split}
\end{equation}
An alternating minimization scheme can be used to solve (\ref{equ:pwls-csc-2}). First, an intermediate reconstructed image $\bm{u}$ can be obtained with a set of fixed feature maps $\{\widetilde{\bm{M}}_i\}$. Thus, the objective function (\ref{equ:pwls-csc-2}) becomes:
\begin{equation}
  \label{equ:pwls-csc-u}
  \begin{split}
    \mathop{arg\ min}_{\bm{u}}
    &(\bm{y}-\bm{Au})^T \bm{\Sigma}^{-1} (\bm{y}-\bm{Au}) \\
    & + \beta\left\|\sum_{i=1}^N \bm{f}_i \ast \widetilde{\bm{M}}_i - \bm{u}\right\|_2^2.
  \end{split}
\end{equation}
With the separable paraboloid surrogate method \cite{Elbakri2002Statistical}, the solution of (\ref{equ:pwls-csc-u}) can be obtained by:
\begin{equation}
  \label{equ:solution-u}
  \begin{split}
    \bm{u}_q^{t+1} = \bm{u}_q^t
     & - \frac{\sum_{p=1}^P\left(\left(1/\sigma_p^2\right)a_{pq}\left(\left[\bm{Au}^t\right]_p - \bm{y}_p\right)\right)}{\sum_{p=1}^P\left(\left(1/\sigma_p^2\right)a_{pq}\sum_{k=1}^Q a_{pk}\right) + \beta} \\
     & - \frac{\beta\left(\left[\sum_{i=1}^N \bm{f}_i \ast \widetilde{\bm{M}}_i\right]_q^t-\bm{u}_q^t\right)}{\sum_{p=1}^P\left(\left(1/\sigma_p^2\right)a_{pq}\sum_{k=1}^Q a_{pk}\right) + \beta},\\
     & q = 1, 2, \dots, Q,
  \end{split}
\end{equation}
where $t = 0, 1, 2, \dots, T$ represents the iteration index and $a_{pq}$ is the element of system matrix $\bm{A}$. The second step is to represent the intermediate result $\bm{u}$ obtained in the last step with the fixed filters $\{\bm{f}_i\}$, which means calculating $\{\bm{M}_i\}$. Since CSC does not provide a good representation of the low-frequency component of the image \cite{Wohlberg2016Convolutional}, only the high-frequency component of $\bm{u}$ is processed by CSC, and the results can be obtained by simply summing the recovered high-frequency component and the low-frequency component. The high-frequency component of the image is precomputed by a high-pass filtering operation. Note that the symbol $\bm{u}$ in the remainder of this section denotes the high frequency component. Then, we have the following optimization problem:
\begin{equation}
  \label{equ:pwls-csc-z}
  \mathop{arg\ min}_{\{\bm{M}_i\}}\frac{1}{2}\left\|\sum_{i=1}^N \bm{f}_i \ast \bm{M}_i - \bm{u}\right\|_2^2 + \lambda \sum_{i=1}^N \left\| \bm{M}_i \right\|_1.
\end{equation}
Let $\bm{F}_i\bm{M}_i = \bm{f}_i \ast \bm{M}_i$, $\bm{F} = \left(\bm{F}_1\ \bm{F}_2\ \cdots \ \bm{F}_N\right)$ and $\bm{M} = \left(\bm{M}_1\ \bm{M}_2\ \cdots \ \bm{M}_N\right)^T$, (\ref{equ:pwls-csc-z}) can be rewritten as:
\begin{equation}
  \label{equ:pwls-csc-z-new}
  \mathop{arg\ min}_{\bm{M}}\frac{1}{2}\left\|\bm{FM} - \bm{u}\right\|_2^2 + \lambda\left\|\bm{M}\right\|_1.
\end{equation}
The alternating Direction Method of Multipliers (ADMM) \cite{Boyd2011Distributed} is adopted to solve (\ref{equ:pwls-csc-z-new}) by introducing a dual variable $\bm{B}$ that is constrained to be equal to $\bm{M}$, leading to the following problem:
\begin{equation}
  \label{equ:pwls-csc-z-admm}
  \mathop{arg\ min}_{\bm{M},\bm{B}}\frac{1}{2}\left\|\bm{FM} - \bm{u}\right\|_2^2 + \lambda\left\|\bm{B}\right\|_1 \ \ s.t. \ \ \bm{M}=\bm{B}.
\end{equation}
$\bm{M}$ and $\bm{B}$ are optimized alternately as follows:
\begin{equation}
  \label{equ:admm-1st}
  \bm{M}^{j+1} = \mathop{arg\ min}_{\bm{M}} \frac{1}{2}\left\|\bm{FM} - \bm{u}\right\|_2^2 + \frac{\rho}{2}\left\|\bm{M}-\bm{B}^j + \bm{C}^j\right\|_2^2,
\end{equation}
\begin{equation}
  \label{equ:admm-2nd}
  \bm{B}^{j+1} = \mathop{arg\ min}_{\bm{B}} \lambda\left\|\bm{B}\right\|_1 + \frac{\rho}{2}\left\|\bm{M}^{j+1}-\bm{B} + \bm{C}^j\right\|_2^2,
\end{equation}
\begin{equation}
  \label{equ:admm-3rd}
  \bm{C}^{j+1} = \bm{C}^j + \bm{M}^{j+1} -\bm{B}^{j+1}.
\end{equation}
(\ref{equ:admm-1st}) can be solved by transforming it into Fourier domain \cite{Bristow2013Fast, Rusu2014Explicit}, which gives us:
\begin{equation}
  \label{equ:dft}
    \mathop{arg\ min}_{\hat{\bm{M}}} \frac{1}{2}\left\|\hat{\bm{F}}\hat{\bm{M}} - \hat{\bm{u}}\right\|_2^2 + \frac{\rho}{2}\left\|\hat{\bm{M}}-\hat{\bm{B}} + \hat{\bm{C}}\right\|_2^2,
\end{equation}
where $\hat{\bm{F}}$, $\hat{\bm{M}}$, $\hat{\bm{u}}$, $\hat{\bm{B}}$ and $\hat{\bm{C}}$ denote the corresponding expressions in the Fourier domain. (\ref{equ:dft}) can be solved with the following closed-form solution:
\begin{equation}
  \label{equ:sol-dft}
  \left(\hat{\bm{F}}^H\hat{\bm{F}} + \rho\bm{I}\right)\hat{\bm{M}} = \hat{\bm{F}}^H\hat{\bm{u}} + \rho\left(\hat{\bm{B}}-\hat{\bm{C}}\right),
\end{equation}
where $\bm{I}$ is the identity matrix. (\ref{equ:sol-dft}) can be computed efficiently with the Sherman Morrison formula \cite{Wohlberg14Efficient}. The closed-form solution of (\ref{equ:admm-2nd}) can be obtained by:
\begin{equation}
  \label{equ:sol-G}
  \bm{B}^{j+1} = S_{\lambda/\rho}\left(\bm{M}^{j+1} + \bm{C}^j\right),
\end{equation}
where $S(\cdot)$ denotes the soft-thresholding function \cite{Bioucas2007New}, which can be calculated as follows:
\begin{equation}
  \label{equ:soft_thresholding}
  S_l\left(\bm{x}\right) = sign\left(\bm{x}\right)max\left(0, \left|\bm{x}\right| - l\right).
\end{equation}
The overall PWLS-CSC algorithm for sparse-view CT reconstruction is summarized in Algorithm 1.
\begin{algorithm}[t]
  \label{algo:pwls-csc}
  \caption{PWLS-CSC}
  \begin{algorithmic}[1]
    \STATE Initialize: $\bm{y}, \bm{A}, \{\bm{f}_i\}, \beta, \lambda, \rho$
    \STATE Initialize: $\bm{u}^0 = \bm{M}^0 = \widetilde{\bm{M}} = \bm{B}^0 = \bm{C}^0 = \bm{0}$
    \REPEAT
    \FOR {$t = 0, 1, 2, \dots, T$}
    \STATE update $\bm{u}^t$ to $\bm{u}^{t+1}$ by (\ref{equ:solution-u})
    \ENDFOR
    \FOR {$j = 0, 1, 2, \dots, J$}
    \STATE update $\bm{M}^j$ to $\bm{M}^{j+1}$ by solving (\ref{equ:admm-1st})
    \STATE update $\bm{B}^j$ to $\bm{B}^{j+1}$ by solving (\ref{equ:admm-2nd})
    \STATE update $\bm{C}^j$ to $\bm{C}^{j+1}$ by (\ref{equ:admm-3rd})
    \ENDFOR
    \STATE $\widetilde{\bm{M}} = \bm{M}^{j+1}$
    \STATE $\bm{M}^0 = \bm{B}^0 = \bm{C}^0 = \bm{0}$
    \STATE $\bm{u}^0 = \bm{u}^{t+1}$
    \UNTIL {stopping criterion is satisfied}
    \STATE Output: $\bm{u}^{t+1}$
  \end{algorithmic}
\end{algorithm}

\subsection{PWLS-CSCGR}
As claimed in \cite{Xu2012Low}, for DL-based methods, the structure loss or new artifacts may appear due to the inaccurate dictionary atoms. Similarly, CSC-based methods will suffer this problem if inaccurate filters exist. The constraint with gradient regularization (isotropic total variation), which is usually used to suppress outliers, is an effective way to overcome this problem \cite{Wohlberg2016Convolutional, Wohlberg2018Convolutional}. Furthermore, as suggested by \cite{Wohlberg2018Convolutional}, imposing the gradient constraint on the feature maps is superior to applying the constraint on the image domain. As a result, to further improve the proposed PWLS-CSC, in this subsection, we propose to utilize the gradient regularization (PWLS-CSCGR) on feature maps as an additional constraint in PWLS-CSC, which yields the following optimization problem:
\begin{equation}
  \label{equ:pwls-cscgr}
  \begin{split}
    &\mathop{arg\ min}_{\bm{u},\{\bm{M}_i\},\{\bm{f}_i\}}
     \frac{1}{2}(\bm{y}-\bm{Au})^T \bm{\Sigma}^{-1} (\bm{y}-\bm{Au}) + \\
     &\beta(\frac{1}{2}\left\|\sum_{i=1}^N \bm{f}_i \ast \bm{M}_i - \bm{u}\right\|_2^2 + \lambda \sum_{i=1}^N \left\| \bm{M}_i \right\|_1 + \\
     &\frac{\tau}{2}\sum_{i = 1}^{N}\left\|\sqrt{(\bm{g}_0 \ast \bm{M}_i)^2 + (\bm{g}_1 \ast \bm{M}_i)^2} \right\|_2^2).
  \end{split}
\end{equation}
where $\bm{g}_0$ and $\bm{g}_1$ are the filters that compute the gradient along the x- and y-axes respectively, and $\tau$ is a hyperparameter. Similar to (\ref{equ:pwls-csc-2}), given the predetermined filters $\{\bm{f}_i\}$, (\ref{equ:pwls-cscgr}) can be decomposed into two optimization problems. One problem is exactly the same as (\ref{equ:pwls-csc-u}), and the other can be given as:
\begin{equation}
  \label{equ:cscgr-new}
  \begin{split}
  \mathop{arg\ min}_{\{\bm{M}_i\}}
  &\frac{1}{2}\left\|\sum_{i=1}^N \bm{f}_i \ast \bm{M}_i - \bm{u}\right\|_2^2 + \lambda \sum_{i=1}^N\left\|\bm{M}_i\right\|_1 + \\
  &\frac{\tau}{2}\sum_{i = 1}^{N}\left\|\sqrt{(\bm{g}_0 \ast \bm{M}_i)^2 + (\bm{g}_1 \ast \bm{M}_i)^2} \right\|_2^2,
\end{split}
\end{equation}
The linear operators $\bm{G}_0$ and $\bm{G}_1$ are defined such that $\bm{G}_l\bm{M}_i = \bm{g}_l \ast \bm{M}_i$ and the gradient term in (\ref{equ:cscgr-new}) can be rewritten as:
\begin{equation}
  \label{equ:grad-term-rewrite}
  \frac{\tau}{2}\sum_{i = 1}^{N}\left(\left\|\bm{G}_0 \bm{M}_i\right\|_2^2 + \left\|\bm{G}_1 \bm{M}_i \right\|_2^2\right).
\end{equation}
By introducing block matrix notation, (\ref{equ:cscgr-new}) can be rewritten as follows:
\begin{equation}
  \label{equ:cscgr-block}
  \begin{split}
  \mathop{arg\ min}_{\bm{M}}
  &\frac{1}{2}\left\|\bm{FM} - \bm{u}\right\|_2^2 + \lambda\left\|\bm{M}\right\|_1 +\\ &\frac{\tau}{2}\left\|\bm{\Phi}_0\bm{M}\right\|_2^2 + \frac{\tau}{2}\left\|\bm{\Phi}_1\bm{M}\right\|_2^2,
\end{split}
\end{equation}
where
\begin{equation}
  \bm{\Phi}_l =
  \begin{pmatrix}
    \bm{G}_l & \bm{0} & \cdots \\
    \bm{0} & \bm{G}_l & \cdots \\
    \vdots & \vdots & \ddots
  \end{pmatrix}.
\end{equation}
To apply ADMM to (\ref{equ:cscgr-block}), we introduce a dual variable $\bm{B}$ and have the following optimization problem:
\begin{equation}
  \label{equ:admm-cscgr-block}
  \begin{split}
  \mathop{arg\ min}_{\bm{M},\bm{B}}
  &\frac{1}{2}\left\|\bm{FM} - \bm{u}\right\|_2^2 + \frac{\tau}{2}\left\|\bm{\Phi}_0\bm{M}\right\|_2^2 + \\ &\frac{\tau}{2}\left\|\bm{\Phi}_1\bm{M}\right\|_2^2 + \lambda\left\|\bm{B}\right\|_1, \ \ s.t. \ \ \bm{M}=\bm{B}.
  \end{split}
\end{equation}
Then we have an iterative scheme for (\ref{equ:admm-cscgr-block}) as follows:
\begin{equation}
  \label{equ:admm-cscgr-1st}
  \begin{split}
  \bm{M}^{j+1} =
  &\mathop{arg\ min}_{\bm{M}} \frac{1}{2}\left\|\bm{FM} - \bm{u}\right\|_2^2 + \frac{\tau}{2}\left\|\bm{\Phi}_0\bm{M}\right\|_2^2 \\
  & + \frac{\tau}{2}\left\|\bm{\Phi}_1\bm{M}\right\|_2^2 + \frac{\rho}{2}\left\|\bm{M}-\bm{B}^j + \bm{C}^j\right\|_2^2,
\end{split}
\end{equation}
\begin{equation}
  \label{equ:admm-cscgr-2nd}
  \bm{B}^{j+1} = \mathop{arg\ min}_{\bm{B}} \lambda\left\|\bm{B}\right\|_1 + \frac{\rho}{2}\left\|\bm{M}^{j+1}-\bm{B} + \bm{C}^j\right\|_2^2,
\end{equation}
\begin{equation}
  \label{equ:admm-cscgr-3rd}
  \bm{C}^{j+1} = \bm{C}^j + \bm{M}^{j+1} -\bm{B}^{j+1}.
\end{equation}
The only difference between solving (\ref{equ:admm-cscgr-block}) and (\ref{equ:pwls-csc-z-admm}) is solving $\bm{M}$. The solution for (\ref{equ:admm-cscgr-1st}) in the Fourier domain is given by:
\begin{equation}
  \label{equ:sol-cscgr-dft}
  \begin{split}
  &\left(\hat{\bm{F}}^H\hat{\bm{F}} + \tau\hat{\bm{\Phi}}_0^H\hat{\bm{\Phi}}_0 + \tau\hat{\bm{\Phi}}_1^H\hat{\bm{\Phi}}_1 + \rho\bm{I}\right)\hat{\bm{M}} \\
  &= \hat{\bm{F}}^H\hat{\bm{u}} + \rho\left(\hat{\bm{B}}-\hat{\bm{C}}\right),
  \end{split}
\end{equation}
that can also be solved by the Sherman Morrison formula \cite{Wohlberg2016Efficient}.
The main steps of the proposed PWLS-CSCGR algorithm are summarized in Algorithm 2.
\begin{algorithm}[t]
  \label{algo:pwls-cscgr}
  \caption{PWLS-CSCGR}
  \begin{algorithmic}[1]
    \STATE Initialize: $\bm{y}, \bm{A}, \{\bm{f}_i\}, \bm{g}_0, \bm{g}_1, \beta, \lambda, \rho, \tau$
    \STATE Initialize: $\bm{u}^0 = \bm{M}^0 = \widetilde{\bm{M}} = \bm{B}^0 = \bm{C}^0 = \bm{0}$
    \REPEAT
    \FOR {$t = 0, 1, 2, \dots, T$}
    \STATE update $\bm{u}^t$ to $\bm{u}^{t+1}$ by (\ref{equ:solution-u})
    \ENDFOR
    \FOR {$j = 0, 1, 2, \dots, J$}
    \STATE update $\bm{M}^j$ to $\bm{M}^{j+1}$ by solving (\ref{equ:admm-cscgr-1st})
    \STATE update $\bm{B}^j$ to $\bm{B}^{j+1}$ by solving (\ref{equ:admm-cscgr-2nd})
    \STATE update $\bm{C}^j$ to $\bm{C}^{j+1}$ by (\ref{equ:admm-cscgr-3rd})
    \ENDFOR
    \STATE $\widetilde{\bm{M}} = \bm{M}^{j+1}$
    \STATE $\bm{M}^0 = \bm{B}^0 = \bm{C}^0 = \bm{0}$
    \STATE $\bm{u}^0 = \bm{u}^{t+1}$
    \UNTIL {stopping criterion is satisfied}
    \STATE Output: $\bm{u}^{t+1}$
  \end{algorithmic}
\end{algorithm}

\section{EXPERIMENTAL RESULTS}
\label{sec:results}
In this section, extensive experiments were conducted to demonstrate the performance of the proposed methods. Two types of data, including simulated and real clinical data, were used in our experiments. All experiments were performed in MATLAB 2017a on a PC (equipped with an AMD Ryzen 5 1600 CPU at 3.2 GHz and 16 GB RAM). To accelerate the algorithm, we performed the CSC and CSCGR phases with a graphics processing unit (GPU) GTX 1080 Ti. All filters and the regularization parameters used in this section were the same if there was no special instruction, including 32 filters with sizes of 10 $\times$ 10, $\beta$ = 0.005, $\lambda$ = 0.005, $\rho$ = 100 $\times\ \lambda$ + 1 and $\tau$ = 0.06. Notably, in our numerical scheme, the computation of convolution was transformed into the frequency domain, which can be treated as the dot product operation, so there was no need to keep the filter size to an odd number. All the parameters were constants except $\rho$, which was self-adaptive according to \cite{He2000Alternating}. The peak signal-to-noise ratio (PSNR), root-mean-square error (RMSE) and structural similarity (SSIM) \cite{Wang2004Image} were utilized to quantitatively evaluate the performance of the methods.

\subsection{Simulated Results}
\label{subsec:clinical}
\begin{figure}[t]
\begin{center}
\setlength{\abovecaptionskip}{0.cm}
\setlength{\belowcaptionskip}{-0.cm}
\includegraphics[width=3.45in]{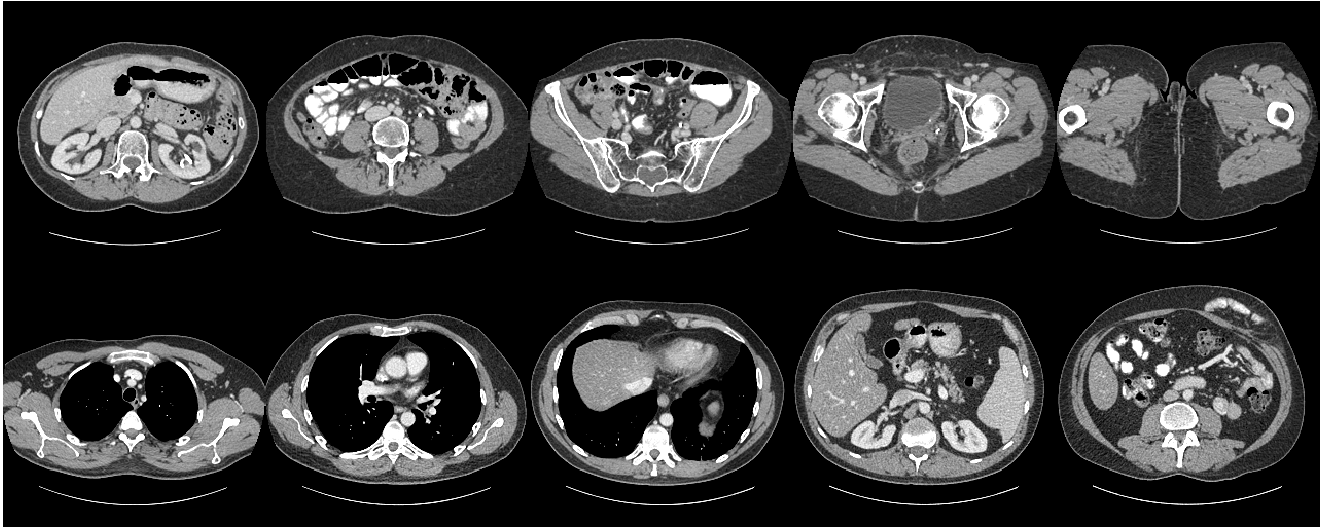}
\caption{Training images in the set. The display window is [-150 250]HU.}
\label{fig:training_image}
\end{center}
\end{figure}

To demonstrate the performance of the proposed methods, a dataset of clinical CT images, authorized by Mayo Clinic and downloaded from "the NIH-AAPM-Mayo Clinic Low Dose CT Grand Challenge," was used \cite{NIH-AAPM-Mayo}. This dataset contains 5,936 1 mm thick full dose CT slices collected from 10 patients. The high-frequency components of the ten images randomly selected from this dataset were used to train the predetermined filters $\{\bm{f}_i\}$. Fig. \ref{fig:training_image} shows the images in the training set. The projection data in fan-beam geometry were simulated by the Siddon's ray-driven algorithm \cite{Siddon1985Fast}. A total of 64 projection views were evenly distributed over $360^o$, and a flat detector with a length of 41.3 cm had 512 bins. The image arrays were 20 $\times$ 20 cm$^2$. The distance from the source to the center of rotation was 40 cm, and the distance from the center of the detector to the center of rotation was 40 cm. The outer iteration number was set to 2000. The inner iteration numbers were set to 20 and 100 for PWLS and CSCGR, respectively. The iteration numbers for PWLS-CSC were the same as those for PWLS-CSCGR.
\setlength{\textfloatsep}{0\normalbaselineskip}

Several state-of-the-art methods were compared with our proposed algorithms, including FBP, PWLS-TGV \cite{Niu2014Sparse}, PWLS-DL \cite{Xu2012Low}, PWLS-ULTRA\cite{Zheng2018PWLS} and LEARN \cite{Chen2018LEARN}.

\subsubsection{Abdominal Case}
\begin{figure*}[t]
\begin{center}
\setlength{\abovecaptionskip}{0.cm}
\setlength{\belowcaptionskip}{-0.cm}
\includegraphics[width=6in]{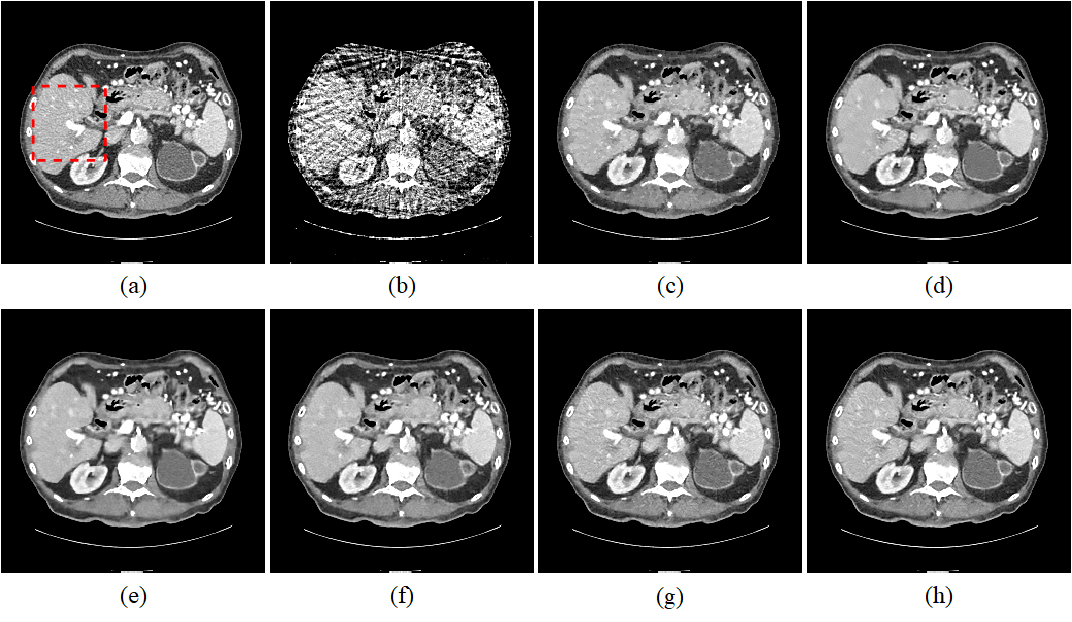}
\caption{Abdominal images reconstructed by the various methods. (a) The reference image versus the images reconstructed by (b) FBP, (c) PWLS-TGV, (d) PWLS-DL, (e) PWLS-ULTRA, (f) LEARN, (g) PWLS-CSC and (h) PWLS-CSCGR. The red box labels a region of interest (ROI) to be zoomed-in. The display window is [-150 250]HU.}
\label{fig:abdominal_image}
\end{center}
\end{figure*}

\begin{figure*}[t]
\begin{center}
\setlength{\abovecaptionskip}{0.cm}
\setlength{\belowcaptionskip}{-0.cm}
\includegraphics[width=6in]{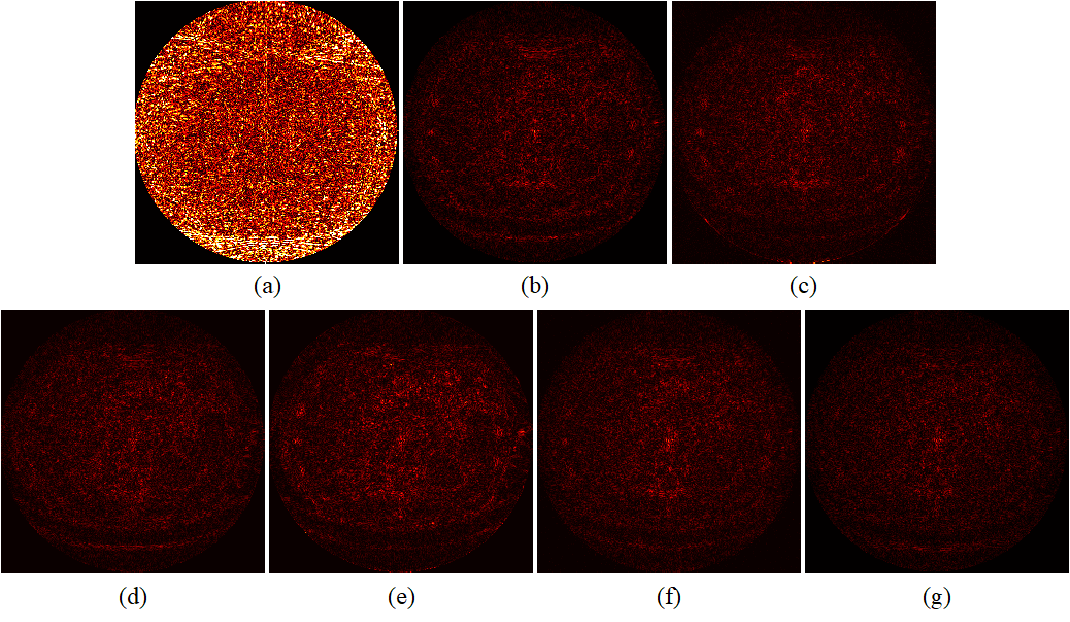}
\caption{The difference images are relative to the original image. Results for (a) FBP, (b) PWLS-TGV, (c) PWLS-DL, (d) PWLS-ULTRA, (e) LEARN, (f) PWLS-CSC and (g) PWLS-CSCGR. The display window is [-1000 -700]HU.}
\label{fig:abdominal_diff}
\end{center}
\end{figure*}

The original abdominal image and the results reconstructed from different methods are shown in Fig. \ref{fig:abdominal_image}. The result of FBP contains severe streak artifacts and is clinically almost useless, while all other algorithms efficiently remove the artifacts. However, in Fig. \ref{fig:abdominal_image}(c), blocky effects are noticed to a certain degree. PWLS-DL effectively avoids the blocky effects in Fig. \ref{fig:abdominal_image}(d), but the result appears smooth in some organs. This effect is because the DL method is patch-based, which usually averages all the overlapped patches to produce the final result. This procedure may efficiently suppress the noise, but some small details may also be smoothened. PWLS-ULTRA suffers the same problem as PWLS-DL. In Fig. \ref{fig:abdominal_image}(f), the recent deep learning-based method LEARN also fails to recover small details. In Fig. \ref{fig:abdominal_image}(g) and (h), the proposed methods maintain more details while the artifacts are removed, especially some contrast-enhanced blood vessels in the liver. The differences between PWLS-CSC and PWLS-CSCGR are almost invisible to visual inspection, but the absolute difference images, which are illustrated in Fig. \ref{fig:abdominal_diff}, show that the PWLS-CSCGR exhibits better performance.
\setlength{\textfloatsep}{0\normalbaselineskip}

\setlength{\textfloatsep}{0\normalbaselineskip}

\begin{figure}[t]
\begin{center}
\setlength{\abovecaptionskip}{0.cm}
\setlength{\belowcaptionskip}{-0.cm}
\includegraphics[width=3.45in]{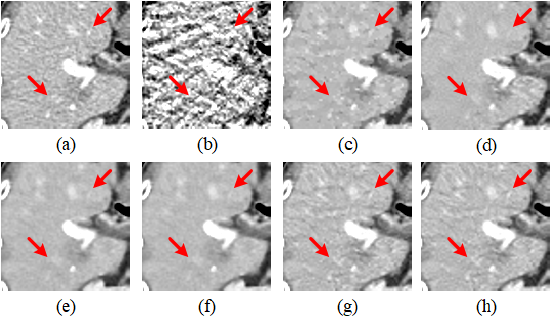}
\caption{Enlarged region of interest (ROI) indicated by the red box in Fig. \ref{fig:abdominal_image}(a). (a) The reference image versus the images reconstructed by (b) FBP, (c) PWLS-TGV, (d) PWLS-DL, (e) PWLS-ULTRA, (f) LEARN, (g) PWLS-CSC and (h) PWLS-CSCGR. The red arrows indicate that two differences can be easily identified by our proposed methods. The display window is [-150 250]HU.}
\label{fig:abdominal_zoom_in}
\end{center}
\end{figure}

To better visualize the details, Fig. \ref{fig:abdominal_zoom_in} shows the results of a magnified region of interest (ROI), which is indicated by the red box in Fig. \ref{fig:abdominal_image}(a). The red arrows indicate several contrast-enhanced blood vessels in the liver, which can only be well identified by our methods. In particular, in Fig. \ref{fig:abdominal_zoom_in}(d)-(f), the liver region is smoothened, and the contrast-enhanced blood vessels are almost lost. In Fig. \ref{fig:abdominal_zoom_in}(g) and (h), the proposed methods preserve most details and have the most coherent visual effect on the reference image, even for the mottle-like structures in the liver.
\setlength{\textfloatsep}{0\normalbaselineskip}

\subsubsection{Thoracic Case}
\begin{figure*}[t]
\begin{center}
\setlength{\abovecaptionskip}{0.cm}
\setlength{\belowcaptionskip}{-0.cm}
\includegraphics[width=6in]{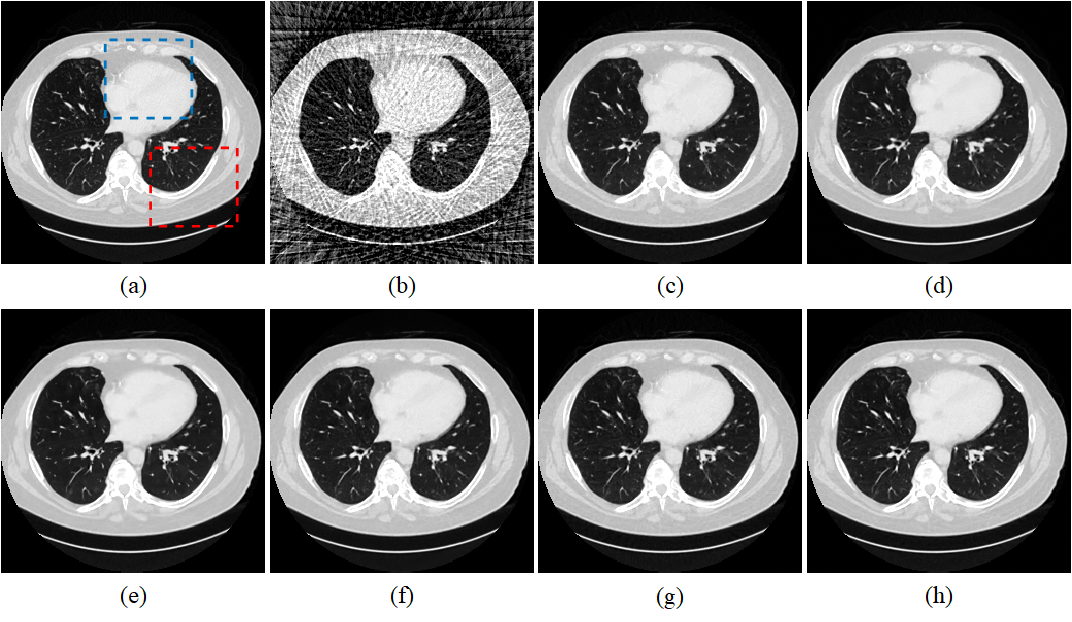}
\caption{Thoracic images reconstructed by the various methods. (a) The reference image versus the images reconstructed by (b) FBP, (c) PWLS-TGV, (d) PWLS-DL, (e) PWLS-ULTRA, (f) LEARN, (g) PWLS-CSC and (h) PWLS-CSCGR. The blue and red boxes label two ROIs to be magnified. The display window is [-1000 250]HU.}
\label{fig:thoracic_image}
\end{center}
\end{figure*}

Fig. \ref{fig:thoracic_image} presents the thoracic images reconstructed by different methods. It can be easily observed that the FBP result is covered by the artifacts, and it is difficult to clearly discriminate the artifacts and blood vessels in the lung. In Fig. \ref{fig:thoracic_image}(c)-(f), most artifacts are suppressed, but the results are smoothened to different degrees in some organs,  and some important details are lost. In Fig. \ref{fig:thoracic_image}(g) and (h), more details in some regions are preserved while the artifacts are effectively removed.

\begin{figure}[t]
\begin{center}
\setlength{\abovecaptionskip}{0.cm}
\setlength{\belowcaptionskip}{-0.cm}
\includegraphics[width=3.45in]{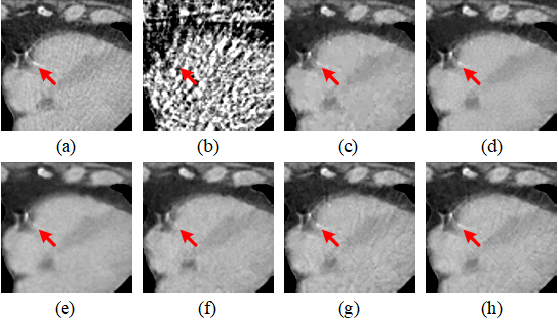}
\caption{Magnified region of interest (ROI) indicated by the blue box in Fig. \ref{fig:thoracic_image}(a). (a) The reference image versus the images reconstructed by (b) FBP, (c) PWLS-TGV, (d) PWLS-DL, (e) PWLS-ULTRA, (f) LEARN, (g) PWLS-CSC and (h) PWLS-CSCGR. The red arrow indicates that the differences can be easily identified by our proposed methods. The display window is [-150 250]HU.}
\label{fig:thoracic_zoom_blue}
\end{center}
\end{figure}

\begin{figure}[t]
\begin{center}
\setlength{\abovecaptionskip}{0.cm}
\setlength{\belowcaptionskip}{-0.cm}
\includegraphics[width=3.45in]{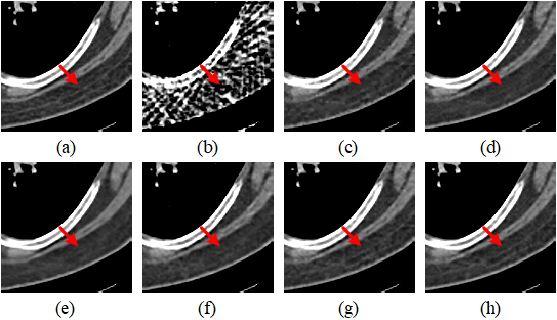}
\caption{Magnified region of interest (ROI) indicated by the red box in Fig. \ref{fig:thoracic_image}(a). (a) The reference image versus the images reconstructed by (b) FBP, (c) PWLS-TGV, (d) PWLS-DL, (e) PWLS-ULTRA, (f) LEARN, (g) PWLS-CSC and (h) PWLS-CSCGR. The red arrow indicates that the differences can be easily identified by our proposed methods. The display window is [-150 250]HU.}
\label{fig:thoracic_zoom_red}
\end{center}
\end{figure}

In particular, two small regions, which are indicated by the blue and red boxes in Fig. \ref{fig:thoracic_image}(a), are enlarged in Fig. \ref{fig:thoracic_zoom_blue} and Fig. \ref{fig:thoracic_zoom_red}. In Fig. \ref{fig:thoracic_zoom_blue}(b), the heart could not be well recognized. PWLS-TGV distorted the boundaries and the structures in the heart in Fig. \ref{fig:thoracic_zoom_blue}(c). As indicated by the red arrow, in Figs. \ref{fig:thoracic_zoom_blue}(d)-(f), the tiny vessels is blurred, and only PWLS-CSC and PWLS-CSCGR satisfactorily recover this structural information. In Fig. \ref{fig:thoracic_zoom_blue}(g), the result of PWLS-CSC introduces ringing artifacts, which are probably caused by inaccurate filters, while PWLS-CSCGR effectively suppresses these artifacts with the aid of gradient regularization. In Fig. \ref{fig:thoracic_zoom_red}, the curve-like structures indicated by the red arrow are clearly visible in the results of PWLS-CSC and PWLS-CSCGR, and all the other methods cannot adequately maintain these structures.
\setlength{\textfloatsep}{0\normalbaselineskip}

\setlength{\textfloatsep}{0\normalbaselineskip}

\setlength{\textfloatsep}{0\normalbaselineskip}

\subsubsection{Quantitative Evaluation}
To demonstrate the robustness of the proposed methods at various sampling rates, experiments with 48, 64 and 80 projection views were performed. Table \ref{table:abdominal} tabulates the quantitative results of the abdominal image, which is shown in Fig. \ref{fig:abdominal_image}(a) and reconstructed from 48, 60 and 80 projection views with different methods. When the number of projection views is set to 48, PWLS-DL and PWLS-ULTRA demonstrate better performance than PWLS-CSC, but PWLS-CSC outperforms them when the number of projection views is set to 64 or 80 and PWLS-CSCGR achieves the best performance in all situations. Table \ref{table:thoracic} displays the quantitative evaluations for the thoracic image shown in Fig. \ref{fig:thoracic_image}(a) and reconstructed from 48, 60 and 80 projection views with different methods. A similar trend is shown in Table \ref{table:thoracic}, in which PWLS-CSCGR obtains the highest scores in all three metrics and PWLS-CSC also outperforms PWLS-TGV and PWLS-DL for these conditions. The plausible reason is that the thoracic image has more small details, which can be better represented by CSC. In summary, the proposed PWLS-CSC and PWLS-CSCGR methods are more robust to different sampling rates than other methods.


\begin{center}
\begin{table*}[t]
\centering
\caption{Quantitative results obtained by different algorithms for the abdominal image.}
\label{table:abdominal}
\begin{tabular*}{500pt}{@{\extracolsep\fill}lccccccccc@{\extracolsep\fill}}
\toprule
Num. of Views & & 48 & & & 64 & & & 80 & \\
 & \textbf{PSNR}  & \textbf{RMSE}  & \textbf{SSIM}& \textbf{PSNR}  & \textbf{RMSE}  & \textbf{SSIM}& \textbf{PSNR}  & \textbf{RMSE}  & \textbf{SSIM} \\
\midrule
FBP & 24.15 &	0.06203	& 0.41375	& 25.35	& 0.05417	& 0.45858	& 26.40 &	0.04785 &	0.54287\\
PWLS-TGV & 41.06 & 0.00885 & 0.96068 & 44.05 & 0.00627 & 0.97697 & 46.91 & 0.00451 & 0.98736\\
PWLS-DL & 41.75 &	0.00817 &	0.96863 &	44.84 &	0.00573 &	0.97975 &	47.65 &	0.00414 & 0.98915\\
PWLS-ULTRA & 41.69 &	0.00824 &	0.96672 &	44.46 &	0.00599 &	0.97791 &	46.02 &	0.00500 & 0.98355\\
LEARN & 40.43 &	0.00953 &	0.95421 &	43.28 &	0.00692 &	0.97254 &	44.74 &	0.00580 & 0.97945\\
PWLS-CSC & 41.36 & 0.00855 & 0.96375 & 44.92 & 0.00568 & 0.98064 & 47.75 & 0.00410 & 0.98911
\\
PWLS-CSCGR & \textbf{42.52} & \textbf{0.00748} & \textbf{0.96919} & \textbf{45.73} & \textbf{0.00517} & \textbf{0.98329} & \textbf{48.38} & \textbf{0.00381} & \textbf{0.99034} \\
\bottomrule
\setlength{\textfloatsep}{0\normalbaselineskip}
\end{tabular*}
\end{table*}
\end{center}

\begin{center}
\begin{table*}[t]
\centering
\caption{Quantitative results obtained by different algorithms for the thoracic image.}
\label{table:thoracic}
\begin{tabular*}{500pt}{@{\extracolsep\fill}lccccccccc@{\extracolsep\fill}}
\toprule
Num. of Views & & 48 & & & 64 & & & 80 & \\
 & \textbf{PSNR}  & \textbf{RMSE}  & \textbf{SSIM}& \textbf{PSNR}  & \textbf{RMSE}  & \textbf{SSIM}& \textbf{PSNR}  & \textbf{RMSE}  & \textbf{SSIM} \\
\midrule
FBP & 21.85 & 0.08079 & 0.36232 & 22.99 & 0.07087 & 0.42335 & 23.9 & 0.0638 & 0.47914 \\
PWLS-TGV & 40.74 & 0.00919 & 0.96781 & 44.43 & 0.00600 & 0.98310 & 48.44	& 0.00379 & 0.99231 \\
PWLS-DL & 39.80 & 0.01023 & 0.96203 & 43.68 & 0.00655 & 0.98145 & 47.21 & 0.00436 & 0.99069\\
PWLS-ULTRA & 42.55 &	0.00745 &	0.97535 &	45.13 &	0.00554 &	0.98399 &	47.04 &	0.00445 & 0.98884\\
LEARN & 39.43 &	0.01068 &	0.95866 &	42.52 &	0.00743 &	0.97716 &	44.29 &	0.00613 & 0.98340\\
PWLS-CSC & 41.55 & 0.00837 & 0.97177 & 45.07 & 0.00558 & 0.98517 & 48.69 & 0.00368 &  0.99269\\
PWLS-CSCGR & \textbf{42.78} & \textbf{0.00726} & \textbf{0.97687} & \textbf{46.07} & \textbf{0.00497}  & \textbf{0.98753} & \textbf{49.54} & \textbf{0.00333} & \textbf{0.99372}\\
\bottomrule 
\setlength{\textfloatsep}{0\normalbaselineskip}
\end{tabular*}
\end{table*}
\end{center}

\subsection{Real Data}
\label{subsec:read_data}
To evaluate the proposed methods in a real clinical configuration, in this subsection, projection data acquired on a UEG Dental CT scanner using a normal-dose protocol of 110 kVp and 13 mA/10 ms per projection were used. Five hundred sampling angles were evenly distributed over 360 degrees. In each projection view, 950 bins were distributed defining a field-of-view (FOV) of 18 $\times$ 18 $\times$ 25 cm$^3$. The reconstructed image is a matrix of 256 $\times$ 256 pixels. The outer iteration number was set to 40, and the inner iteration numbers for PWLS and CSCGR were set to 2 and 100, respectively. The iteration numbers for PWLS-CSC are the same as those for PWLS-CSCGR.

\begin{figure*}[t]
\begin{center}
\setlength{\abovecaptionskip}{0.cm}
\setlength{\belowcaptionskip}{-0.cm}
\includegraphics[width=6in]{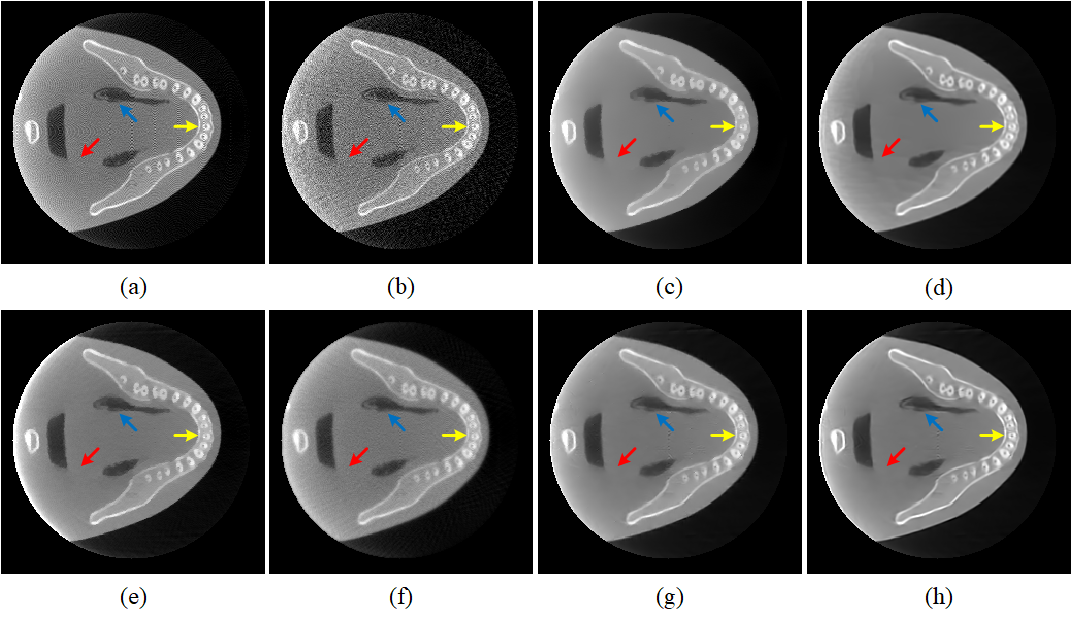}
\caption{Oral images reconstructed by the various methods. (a) The ART result with a full-dose versus the images reconstructed by (b) FBP, (c) PWLS-TGV, (d) PWLS-DL, (e) PWLS-ULTRA, (f) LEARN, (g) PWLS-CSC and (h) PWLS-CSCGR. The arrows indicate that the differences can be easily identified by our proposed methods. The display window is [-1000 200]HU.}
\label{fig:teeth}
\end{center}
\end{figure*}

Fig. \ref{fig:teeth} shows the reference image reconstructed by ART with full-sampled (500 views) projection data and the results reconstructed by various methods with downsampled projection data with 125 views. It can be observed that due to the limitation of hardware, noticeable noise and artifacts still exist in the full-sampled result in Fig. \ref{fig:teeth}(a). In Fig \ref{fig:teeth}(b), FBP suffers from severe artifacts that fulfill every area of the mouth. In Fig. \ref{fig:teeth}(c), PWLS-TGV completely removes most of the artifacts, but the blocky effects at the edges of tissue, which are indicated by the blue and yellow arrows, are obvious. PWLS-DL efficiently suppresses the artifacts shown in the reference image. However, some artifacts caused by patch aggregation can be observed in Fig. \ref{fig:teeth}(d), which are indicated by the red arrow, and the details of the teeth are blurred. In Fig. \ref{fig:teeth}(e) and (f), PWLS-ULTRA and LEARN cannot remove all the artifacts, and the details of teeth are blurred. In Fig. \ref{fig:teeth}(g) and (h), PWLS-CSC and PWLS-CSCGR efficiently eliminate the noise and artifacts and avoid the side effect caused by patch aggregation in PWLS-DL, and PWLS-CSCGR clearly preserves the details of teeth.
\setlength{\textfloatsep}{0\normalbaselineskip}

\subsection{Effect of Filter Parameters}
In this subsection, we evaluate the impact of filter parameters on PWLS-CSCGR, including the number of filters, the filter sizes, the number of training samples and the different types of training sets.

\subsubsection{Number of Filters}
To evaluate the impact of the number of filters, experiments were performed with $N$ = 4, 8, 16, 24, 32, 64 and 150 on the selected images in subsection \ref{subsec:clinical}. The quantitative results are summarized in Table \ref{table:num_of_filters}. It can be seen that the metrics for both images rise at first and then begin to stabilize when $N$ continues to increase. This phenomenon may rely on the fact that when $N$ is small, the performance can be improved significantly by adding more filters due to the increasing precision of representation, but when $N$ is large enough, the benefit will also decay rapidly.  Based on this observation, to balance the performance and computational time, the number of filters was set to 32 in this paper.

\begin{center}
\begin{table*}[t]
\centering
\caption{Quantitative results associated with the different number of filters.}
\label{table:num_of_filters}
\begin{tabular*}{400pt}{@{\extracolsep\fill}lccccccccc@{\extracolsep\fill}}
\toprule
Image & & Abdominal & & & & Thoracic & \\
Num. of Filters & \textbf{PSNR}  & \textbf{RMSE}  & \textbf{SSIM} & Time & \textbf{PSNR}  & \textbf{RMSE}  & \textbf{SSIM} & Time\\
\midrule
4  & 44.77 &	0.00577 &	0.97990	&	722 s & 45.00 &	0.00562 &	0.98450	&	788 s\\
8 & 45.40 &	0.00537 & 0.98205	&	776 s	&	45.74	& 0.00517 &	0.98654	&	695 s\\
16 & 45.54 &	0.00528	& 0.98262	&	745 s	&	46.01	& 0.00500	& 0.98737	&	763 s\\
24 & 45.61 & 0.00524 & 0.98288	&	787 s & 46.07 & 0.00497 &	0.98753	&	917 s\\
32  & \textbf{45.73} & \textbf{0.00517} & 0.98329	&	970 s & 46.07 & 0.00497 & 0.98753	&	985 s\\
64 & 45.70 & 0.00519 & \textbf{0.98334}	&	1069 s & \textbf{46.18} & \textbf{0.00491}  & \textbf{0.98779}	&	1371 s \\
150 & 45.65	& 0.00522	& 0.98308	& 2087 s	&	46.03	& 0.00500	& 0.98752	&	2418 s \\
\bottomrule
\setlength{\textfloatsep}{0\normalbaselineskip}
\end{tabular*}
\end{table*}
\end{center}

\subsubsection{Filter Sizes}
\begin{figure}[t]
\begin{center}
\setlength{\abovecaptionskip}{0.cm}
\setlength{\belowcaptionskip}{-0.cm}
\includegraphics[width=3.45in]{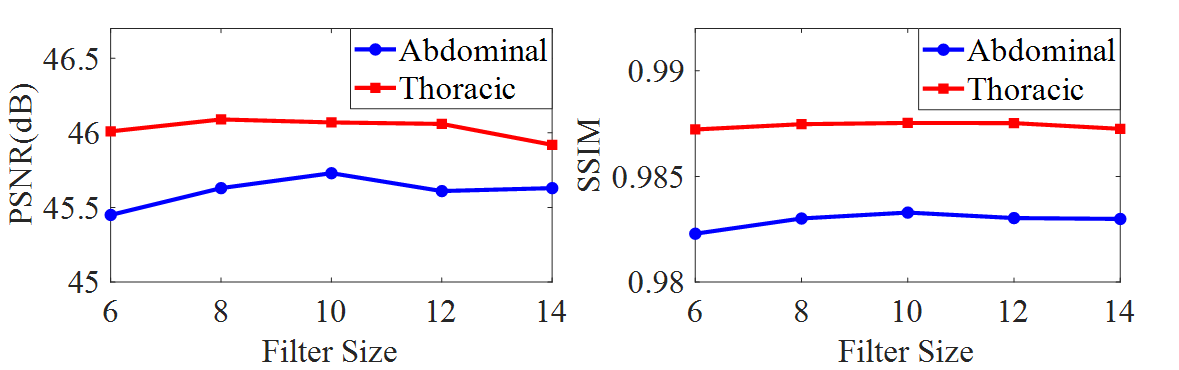}
\caption{Performance with respect to the different filter sizes.}
\label{fig:filter_sizes}
\end{center}
\end{figure}

To sense the impact of filter sizes, different sizes including 6, 8, 10, 12 and 14 were tested. The results are given in Fig. \ref{fig:filter_sizes}. The results show that the performance of our method is not quite sensitive to the filter size and that there are no distinct differences for different sizes. As a result, we simply set filter size to 10, which is approximately the maximum in Fig. \ref{fig:filter_sizes}.
\setlength{\textfloatsep}{0\normalbaselineskip}

\subsubsection{Number of Training Samples}
\label{subsec:num_of_samples}
In this subsection, different numbers of training samples are used to validate the impact of the size of the training set. All images used were randomly selected from the Mayo dataset. The quantitative results are given in Table \ref{table:num_of_samples}. Notably, as the number of training samples increases, the improvement in the performance is small, which can be treated as the evidence that for the current number of filters, the performance is robust to the size of the training set.
\begin{center}
\begin{table*}[t]
\centering
\caption{Quantitative results associated with the different numbers of training samples.}
\label{table:num_of_samples}
\begin{tabular*}{400pt}{@{\extracolsep\fill}lccccccccc@{\extracolsep\fill}}
\toprule
Image & & Abdominal & & & Thoracic & \\
Num. of Samples & \textbf{PSNR}  & \textbf{RMSE}  & \textbf{SSIM} & \textbf{PSNR}  & \textbf{RMSE}  & \textbf{SSIM} \\
\midrule
1	& 45.43	& 0.00535	& 0.98236	& 45.48	& 0.00532	& 0.98595 \\
3	& 45.69	& 0.00519	& 0.98313	& 45.94 &	0.00504	& 0.98717 \\
5	& 45.68	& 0.00520 & 0.98317 & 45.94 & 0.00504 & 0.98727 \\
10 & \textbf{45.73} & 0.\textbf{00517} & \textbf{0.98329} & 46.07 & 0.00497 & 0.98753 \\
50 & 45.68	& 0.00520 & 0.98313 &  \textbf{46.12} & \textbf{0.00494} & \textbf{0.98764} \\
\bottomrule
\end{tabular*}
\end{table*}
\end{center}

\subsubsection{Filter Trained with Different Training Sets}
\label{subsec:fbp}
\begin{figure}[t]
\begin{center}
\setlength{\abovecaptionskip}{0.cm}
\setlength{\belowcaptionskip}{-0.cm}
\includegraphics[width=3.45in]{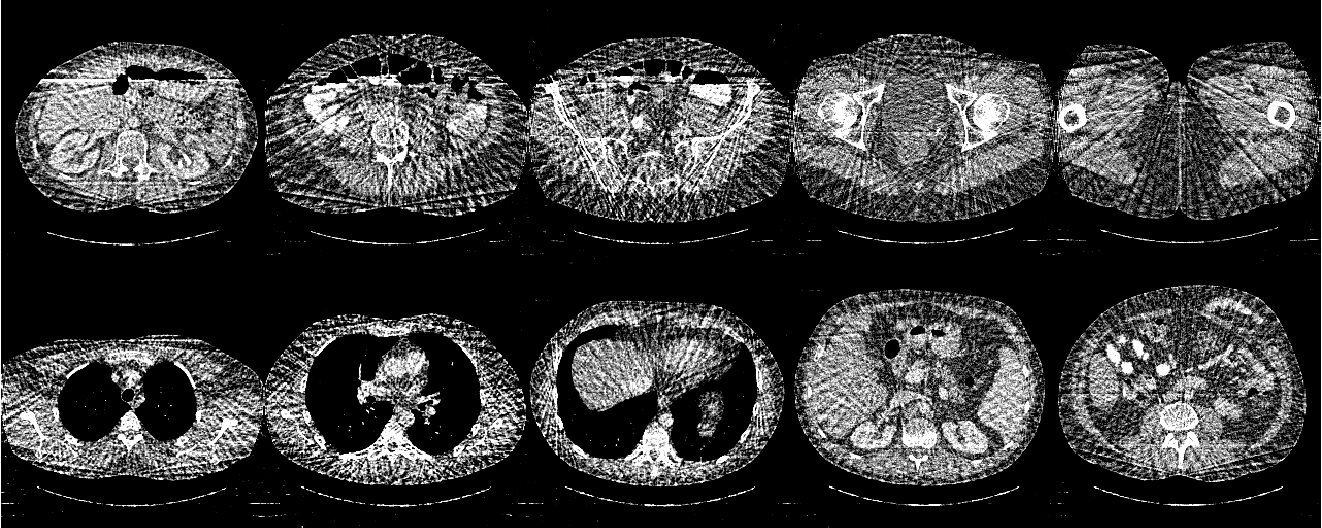}
\caption{Training set with undersampled CT images reconstructed by FBP.}
\label{fig:training_fbp}
\end{center}
\end{figure}

\begin{figure}[t]
\begin{center}
\setlength{\abovecaptionskip}{0.cm}
\setlength{\belowcaptionskip}{-0.cm}
\includegraphics[width=3.45in]{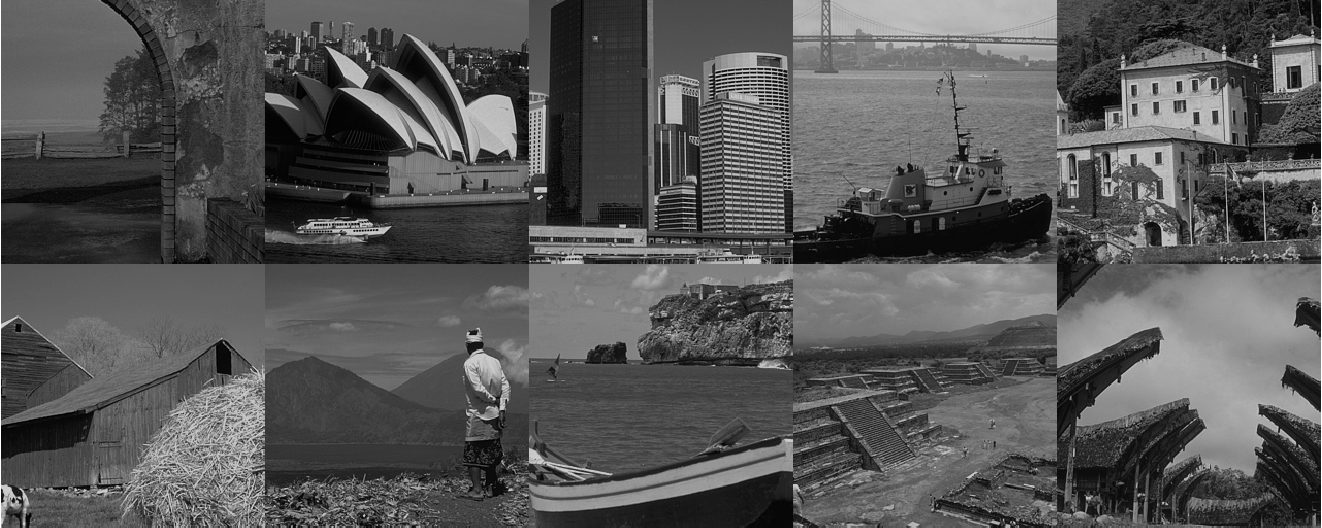}
\caption{Training set with natural images.}
\label{fig:training_natural}
\end{center}
\end{figure}

\begin{figure}[t]
\begin{center}
\setlength{\abovecaptionskip}{0.cm}
\setlength{\belowcaptionskip}{-0.cm}
\includegraphics[width=3.45in]{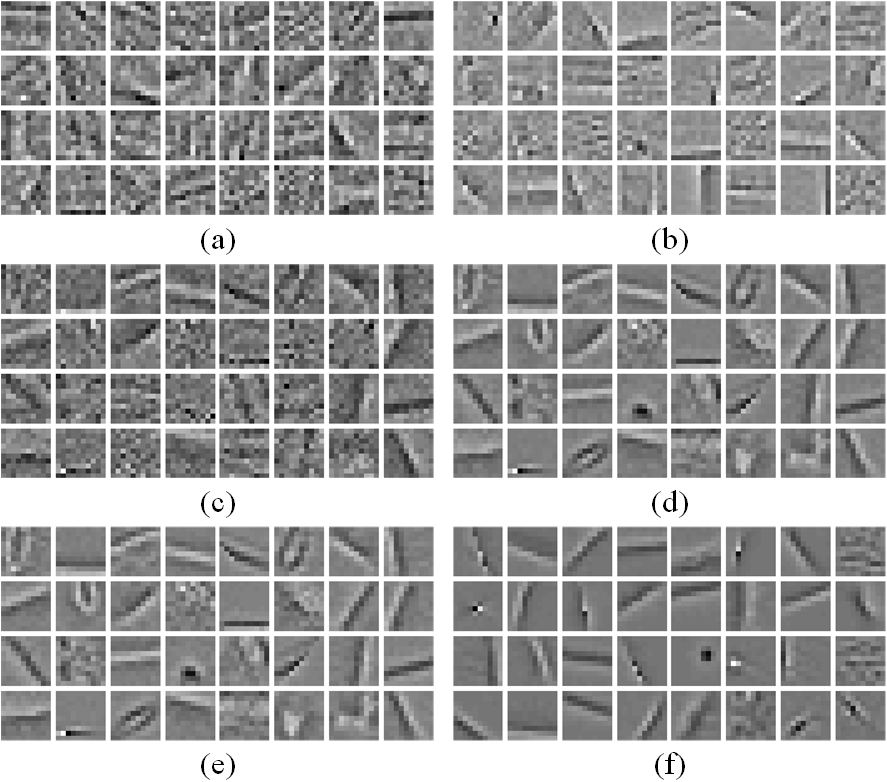}
\caption{Filters trained with (a) undersampled CT images, (b) natural images, (c)-(e) intermediate images (first iteration, 250-$th$ iteration and last iteration) and (f) full-sampled CT images.}
\label{fig:filters}
\end{center}
\end{figure}

\begin{figure*}[t]
\begin{center}
\setlength{\abovecaptionskip}{0.cm}
\setlength{\belowcaptionskip}{-0.cm}
\includegraphics[width=7in]{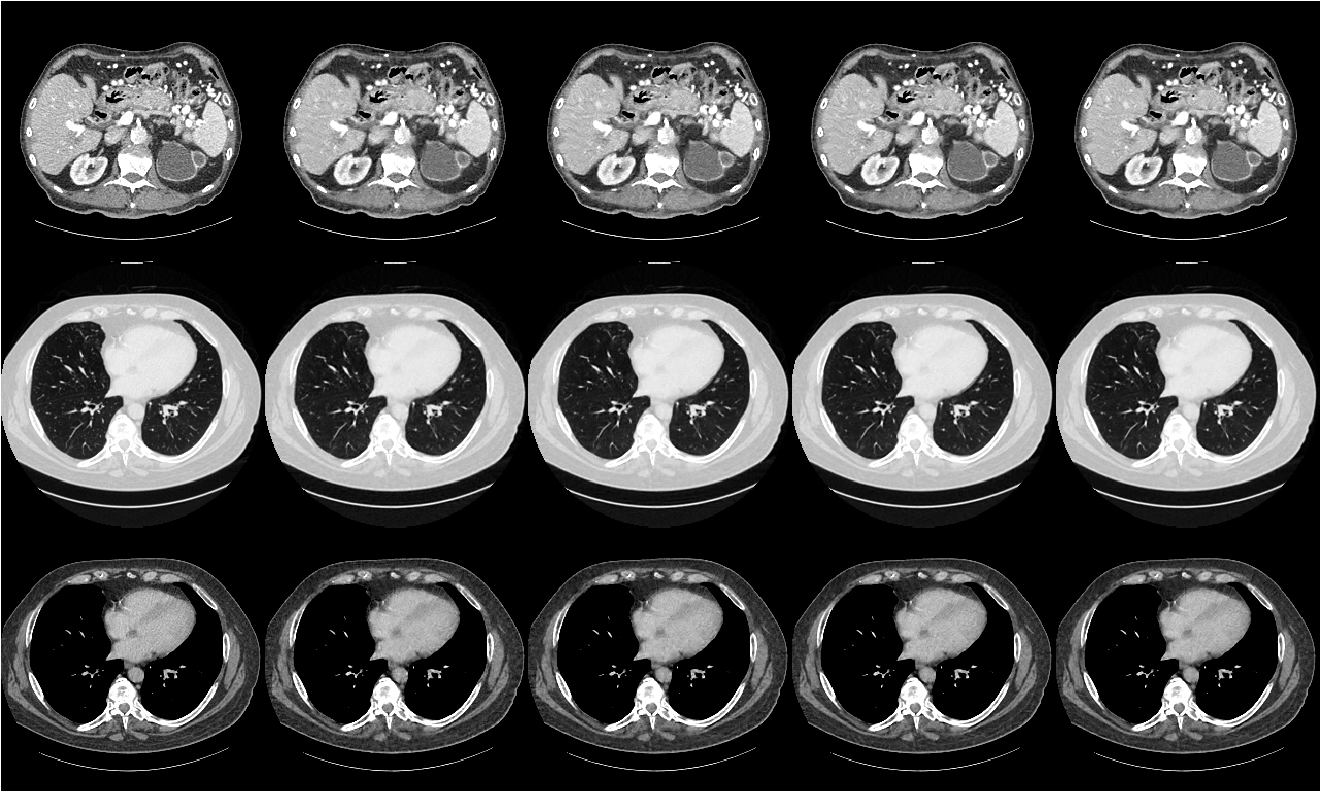}
\caption{Results reconstructed by filters trained with different training sets. The images shown in the second and third rows are the same images with different display windows. (1) The first column gives reference images, (2) the second column shows the results reconstructed by filters trained with undersampled CT images reconstructed by FBP, (3) the third column provides the results reconstructed by filters trained with natural images, (4) the fourth column depicts the results reconstructed by adaptive filters, and (5) the fifth column gives the results reconstructed by filters trained with external full-sampled CT images. The display windows for the first and third rows are [-150 250]HU. The display window for the second row is [-1000 250]HU.}
\label{fig:result_diff_filters}
\end{center}
\end{figure*}

In this subsection, we consider different types of images to form the training set. Three datasets, including undersampled CT images, natural images and the intermediate images in the reconstruction procedure, were tested. Fig. \ref{fig:training_fbp} shows the undersampled CT images reconstructed by FBP. Fig. \ref{fig:training_natural} shows the natural image dataset. The undersampled CT images and natural images were used to train predetermined filters. The adaptive filters are trained with the intermediate images in the reconstruction procedure and updated in each iteration, which is similar to the adaptive dictionary learning in \cite{Xu2012Low}. Thirty-two filters with a size of 10 $\times$ 10 are trained with different training sets, and the filters are shown in Fig. \ref{fig:filters} (the adaptive filters are trained with the intermediate result of the abdominal image). It can be observed that the learned adaptive filters become more sharper and cleaner in Fig. \ref{fig:filters}(c)-(e), and the two pretrained filters set in Fig. \ref{fig:filters}(a) and (b) also capture the main structures although the filters look noisy. Moreover, in Fig. \ref{fig:result_diff_filters}, the reconstruction results with different filters trained with various image sets appear to be visually similar, and it can be seen that the quantitative results shown in Table \ref{table:diff_training_set} are also close. Notably, although the results reconstructed by the adaptive filters achieve high scores, it is time consuming to update the filters in each iteration. In summary, our algorithm is robust to the choice of filters (either predetermined or adaptively learned).
\setlength{\textfloatsep}{0\normalbaselineskip}

\setlength{\textfloatsep}{0\normalbaselineskip}

\setlength{\textfloatsep}{0\normalbaselineskip}

\setlength{\textfloatsep}{0\normalbaselineskip}

\begin{center}
  \begin{table*}[t]
    \centering
    \caption{Quantitative results associated with the different training sets.}
    \label{table:diff_training_set}
      \begin{tabular*}{400pt}{@{\extracolsep\fill}lccccccccc@{\extracolsep\fill}}
      \toprule
      Image & & Abdominal & & & Thoracic & \\
      Training Set & \textbf{PSNR}  & \textbf{RMSE}  & \textbf{SSIM} & \textbf{PSNR}  & \textbf{RMSE}  & \textbf{SSIM} \\
      \midrule
      Undersampled CT Images	& 44.12	& 0.00622	& 0.97759	& 43.73	& 0.00651	& 0.98006 \\
      Natural Images	& 44.71	& 0.00581	& 0.97974	& 44.65	& 0.00586	& 0.98329 \\
      Intermediate Images	& 45.48	& 0.00532	& 0.98237	& 45.65	& 0.00522	& 0.98655 \\
      Full-Sampled CT Images	& \textbf{45.73} & \textbf{0.00517} & \textbf{0.98329} & \textbf{46.07} & \textbf{0.00497}  & \textbf{0.98753} \\
      \bottomrule
    \end{tabular*}
  \end{table*}
\end{center}

\subsection{Convergence analysis and the Effect of Regularization Parameters}
\subsubsection{Convergence analysis}
\begin{figure}[t]
  \begin{center}
  \setlength{\abovecaptionskip}{0.cm}
  \setlength{\belowcaptionskip}{-0.cm}
  \includegraphics[width=3.45in]{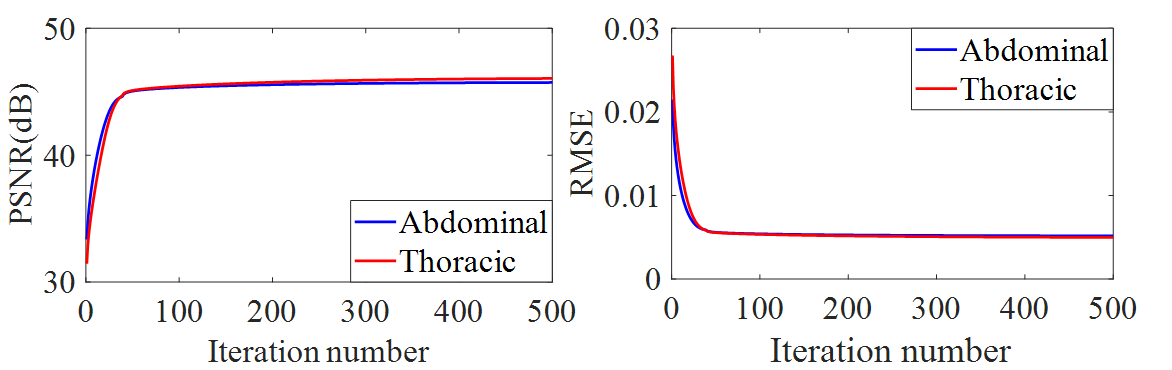}
  \caption{Convergence curves.}
  \label{fig:converges}
  \end{center}
  \end{figure}

Fig. \ref{fig:converges} shows the PSNR and RMSE versus the iteration steps for the PWLS-CSCGR. It can be seen that with an increase in the iteration number, both PSNR and RMSE curves converge to a stable position. These observations demonstrate that the proposed numerical scheme can efficiently optimize the energy functions to a satisfactory solution.

\subsubsection{Effect of $\beta$}
\begin{figure}[t]
\begin{center}
\setlength{\abovecaptionskip}{0.cm}
\setlength{\belowcaptionskip}{-0.cm}
\includegraphics[width=3.45in]{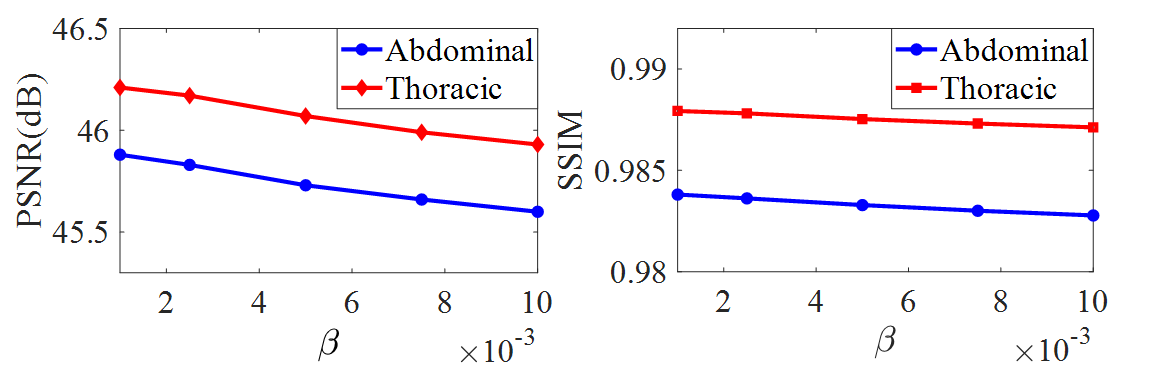}
\caption{Performance with respect to $\beta$.}
\label{fig:alpha}
\end{center}
\end{figure}

To investigate the sensitivity of $\beta$, experiments were performed for various $\beta$ values. The results are presented in Fig. \ref{fig:alpha}. The PSNR and SSIM decrease with increasing $\beta$. However, according to our experiments, the smaller $\beta$ is, the slower the algorithm converges. When $\beta$ is less than or equal to 0.0025, the computational cost is twice or more when $\beta$ is equal to 0.005. Therefore, we set $\beta$ to be 0.005 in our algorithms.

\subsubsection{Effect of $\lambda$}
\begin{figure}[t]
\begin{center}
\setlength{\abovecaptionskip}{0.cm}
\setlength{\belowcaptionskip}{-0.cm}
\includegraphics[width=3.45in]{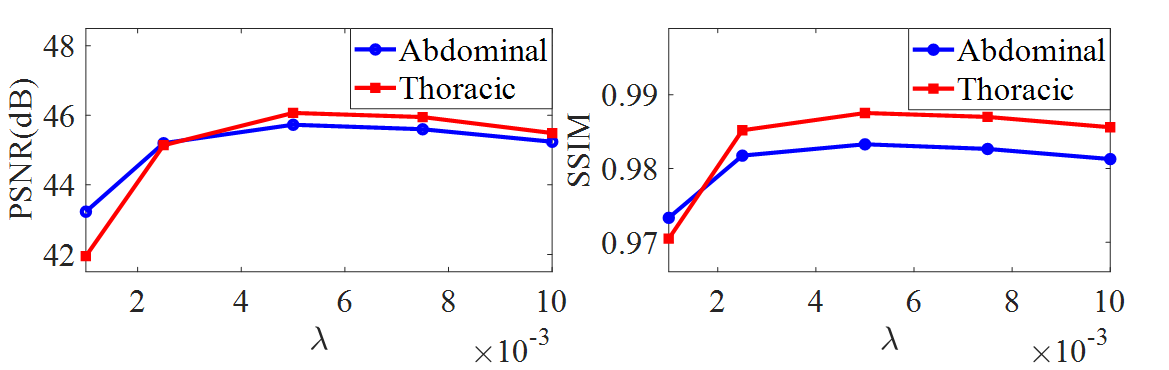}
\caption{Performance with respect to $\lambda$.}
\label{fig:lambda}
\end{center}
\end{figure}

To investigate the sensitivity of our method to $\lambda$, experiments were performed for various $\lambda$ values. The results are shown in Fig. \ref{fig:lambda}. It can be observed that PSNR and SSIM for both images demonstrate similar trends. The values of PSNR and SSIM rise rapidly with an increase in $\lambda$ and reach a peak when $\lambda$ is equal to 0.005. PSNR and SSIM begin to slowly decrease when $\lambda$ is larger than 0.005. Thus, $\lambda$ was set to 0.005 in this paper.

\subsubsection{Effect of $\tau$}
\begin{figure}[t]
\begin{center}
\setlength{\abovecaptionskip}{0.cm}
\setlength{\belowcaptionskip}{-0.cm}
\includegraphics[width=3.45in]{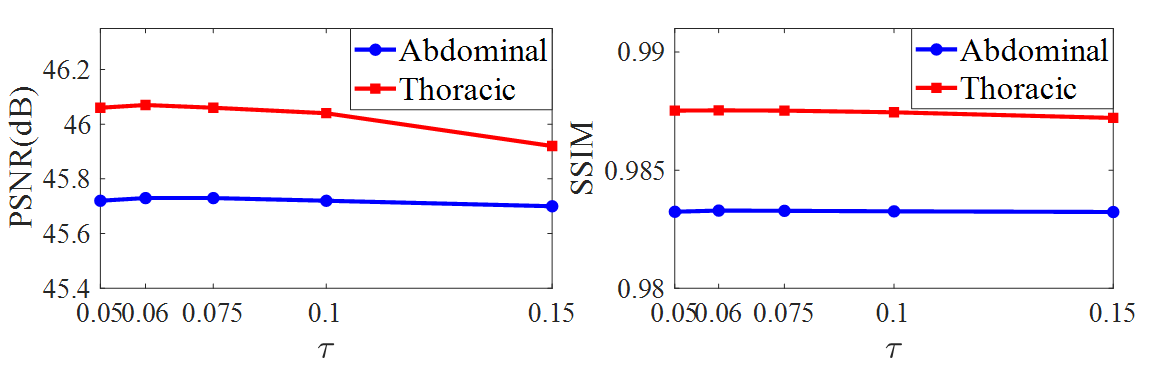}
\caption{Performance with respect to $\tau$.}
\label{fig:mu}
\end{center}
\end{figure}

To explore the impact of $\tau$, experiments were performed for various $\tau$ values. The results are given in Fig. \ref{fig:mu}. It is obvious that PSNR and SSIM change slowly with $\tau$, which means that PWLS-CSCGR is not sensitive to $\tau$. According to the experimental results, $\tau$ was set to 0.06 in our experiments.

\section{DISCUSSION AND CONCLUSION}
\label{sec:discs and concs}
With the development of CSC in recent years, CSC has been proven useful in many imaging problems, including super-resolution, image fusion, image decomposition and so on. Instead of dividing an image into overlapped patches, CSC directly works on the whole image, which maintains more details and avoids artifacts caused by patch aggregation. In this paper, we propose two methods based on CSC. The basic version introduces CSC into the PWLS reconstruction framework. To further improve the performance and preserve more structural information, gradient regularization on feature maps is imposed into the basic version. Qualitative and quantitative results demonstrate the merits of our methods.

In the experiments of Section \ref{subsec:clinical} and \ref{subsec:read_data}, the filters and parameters were the same, showing the generalization of the proposed methods and that there is no need to adjust the filters or the parameters patient by patient. We also examined the impacts of filters on our method. The experimental results show that PWLS-CSCGR can work well even with only four filters. PWLS-CSCGR is also robust to the training set or even without the training set and can be treated as an unsupervised learning method.

Importantly, another issue is the computational time. The main cost of our methods depends on two parts: training the filters and the reconstruction. Training 32 filters with 10 images costs 85 s of GPU. Because this operation is offline and there is no need for a large training set, this part will not be the main problem. On the other hand, the reconstruction is time-consuming. Although our methods have a similar heavy computational burden to PWLS-DL, several techniques, including parallel computing and advanced optimization methods, can be applied for acceleration.

One of the most important deep learning models is CNN, which is also based on the convolution operator. For CSC, a signal can be represented by a summation of convolutions between a set of filters and the corresponding feature maps, and the key point is to calculate the feature maps with certain (predetermined or adaptive) filters. CNN trains the cascaded filters to convolve with the inputs. Furthermore, current CNN-based methods still lack theoretical proof. Most deep learning methods are data-driven, and the results cannot be guaranteed without sufficient training data. However, CSC, as an unsupervised learning method, has a strict mathematical proof. This method is robust to the number of training samples (as shown in Sec. IV-C.3 and IV-C.4) and even without training data. On the other hand, the same groups analyzed the relationship between the CSC and CNN methods in \cite{Papyan2017Convolutional, Papyan2018Theoretical} and found that assuming that our signals originate from the multi-layer CSC model, the layered-thresholding pursuit algorithm for decomposing a given measurement vector $Y$ completely equals the forward propagation in CNNs. This interesting finding provides a new way to explore the interpretability of deep learning.

In conclusion, inspired by successful applications of CSC in the field of signal processing, we explored the potential of this method incorporating a PWLS image reconstruction framework, resulting in two novel algorithms referred to as PWLS-CSC and PWLS-CSCGR. We evaluated the proposed algorithms with simulated and real data. In the experimental results, our methods have been shown to be competitive with several state-of-art methods. The robustness of our methods was also investigated by extensive analysis with experimental configurations. In our future work, we will extend our methods to other CT imaging topics, such as metal artifact reduction and LDCT. Furthermore, the combination with deep learning-based methods is also an interesting direction.

\bibliographystyle{ieeetr}
\bibliography{cite}

\end{document}